\documentclass[journal,10pt]{IEEEtran}

\ifCLASSINFOpdf

\else

\fi

\makeatletter
\def\ifundefined{\@ifundefined}
\makeatother

\usepackage{setspace}
\usepackage{algcompatible}
\usepackage{algorithmicx}
\usepackage{algpseudocode}
\usepackage{algorithm}
\usepackage{multirow}

\usepackage{url}

\usepackage{srcltx}
\usepackage{amsmath}
\usepackage{amssymb}
\usepackage{graphics}
\usepackage{graphicx}
\usepackage{cite}
\usepackage{psfrag}
\usepackage{epsfig}
\usepackage{subfigure}
\usepackage{amsthm}
\usepackage{thmtools}
\usepackage{color}
\usepackage[justification=centering]{caption}
\usepackage{refstyle}
\usepackage{amsmath}
\usepackage{amsthm}
\usepackage[justification=justified,singlelinecheck=false]{caption}

\usepackage{lipsum}

\usepackage{tabularx}
\usepackage{caption}

\newtheorem{theo}{Theorem}
\newtheorem{cor}{Corollary}

\newtheorem{lem}{Lemma}

\usepackage{array,booktabs,arydshln,xcolor}

\usepackage[T1]{fontenc}      
\usepackage[utf8]{inputenc}   
\usepackage{caption}          
\usepackage{mathtools}        
\usepackage{%
   multirow,                   
   tabu                        
}
\usepackage{xcolor}           
\usepackage{siunitx}
\usepackage{float}
\usepackage{arydshln}
\usepackage{booktabs}
\usepackage{lipsum}
\usepackage{makecell}
\setcellgapes{4pt}

\usepackage{mathtools}

\usepackage{adjustbox}
\usepackage{graphicx}

\usepackage{setspace}

\usepackage{enumitem}

\usepackage[justification=centering]{caption}

\begin{document}

\title{Template for IEEE TVT \LaTeX\ Submission}

\title{\huge Local Voting Games for Misbehavior Detection in \\ VANETs in Presence of Uncertainty}



\author{Ali Behfarnia, ~\IEEEmembership{Student Member,~IEEE,} and Ali Eslami, ~\IEEEmembership{Member,~IEEE}

\vspace{-0.15 in}


\thanks{A. Behfarnia and A. Eslami are with the Department of Electrical Engineering and Computer Science, Wichita State University, Wichita, KS, USA ~(emails: axbehfarnia@shockers.wichita.edu, ali.eslami@wichita.edu).}
}

{}

\maketitle

\pagenumbering{gobble}

\begin{abstract}

Cooperation between neighboring vehicles is an effective solution to the problem of malicious node identification in vehicular ad hoc networks (VANETs). However, the outcome is subject to nodes' beliefs and reactions in the collaboration. In this paper, a plain game-theoretic approach that captures the uncertainty of nodes about their monitoring systems, the type of their neighboring nodes, and the outcome of the cooperation is proposed. In particular, one stage of a local voting-based scheme (game) for identifying a target node is developed using a Bayesian game.
In this context, incentives are offered in expected utilities of nodes in order to promote cooperation in the network.
The proposed model is then analyzed to obtain equilibrium points, ensuring that no node can improve its utility by changing its strategy. Finally, the behavior of malicious and benign nodes is studied by extensive simulation results. Specifically, it is shown how the existing uncertainties and the designed incentives impact the strategies of the players and, consequently, the correct target-node identification. 

\end{abstract}

\begin{IEEEkeywords}
Misbehavior detection, local voting-based scheme, game theory, uncertainty, VANETs.
\end{IEEEkeywords}

\IEEEpeerreviewmaketitle

\vspace{-0.1 in}

\section{Introduction}
The high priority of security in intelligent transportation systems has led many researchers to identify security gaps in modern vehicles \cite{Petit15potential,Park17cyber, Survey18}. An important challenge is to detect malicious nodes in vehicular networks, where connections are short-lived (ephemeral), and centrally managed stations are (sometimes) absent. In such transitory distributed networks, quick cooperation among neighboring nodes can provide effective solutions. However, nodes are usually selfish and reluctant to cooperate for no benefit. In addition, each node has some inherent uncertainties in a collaboration, including the type of participants, the accuracy of its own components (e.g., detection system), and attainable outcomes, all of which affect the node’s decision about whether to participate. Therefore, it is crucial to provide incentives according to different reactions of nodes under uncertainty to achieve malicious node detection.

Scholars have realized the effect of misbehaving nodes in the network, and put forward many control and security schemes to mitigate their impact \cite{Raya08revoc, bil10opt, ghan18, kim16, Abass17evol, farshad, masdari17}.
Revocation process is an effective approach for malicious nodes detections that captures the dynamic nature of vehicular ad hoc networks (VANETs) \cite{Raya08revoc, bil10opt}. 
In this process, a benign node is assumed to detect (or get suspicious of) a malicious node and broadcasts its identification (ID) as a target (or an accused) node. 
  Then, other benign neighbors run the voting approach to discredit the target node, while considering their own best interests.   
In this line of work, Kim \cite{kim16} developed a weighted voting-based decision scheme on the basis of cluster architecture to discredit malicious nodes in mobile ad hoc networks. Alabdel et al. \cite{Abass17evol} proposed an evolutionary game model in which all benign nodes take part in the voting game, focusing on unsuccessful revocation and over-reacted revocation decisions. Masdari \cite{masdari17} introduced a collaborative false accusation approach to stop wrong accusations in the network.

Despite valuable efforts in the literature, the study of incentive-based voting games that capture the inherent uncertainties of nodes for malicious-node identification is still incomplete. In this regard, some points should be emphasized.
First, the type of target node could be either malicious or benign, because every node (including malicious nodes) can accuse the others. 
A node can use a detection system to monitor its neighbors, but the accuracy and cost of monitoring should be counted in the game.
Second, benign nodes are uncertain about the strategy of malicious nodes. For instance, a malicious node might intentionally not attack a benign node in order to obtain its support during a voting game. 
Third, incentives should only encourage knowledgeable nodes (i.e., nodes that have already monitored the target node) in cooperating. 
Otherwise, the incentives will lead to many random votes in the game, which might spoil the result of cooperation.
Fourth, both benign nodes and malicious nodes can take part in the voting game. This implies that a benign node cannot rely solely on others' votes, owing to misleading votes from malicious nodes. Finally, the cost of group in the game (a.k.a. social cost) should be designed based on nodes' contributions and their uncertainties about the results. For example, a cooperative node should be punished less than an abstaining node when the collaboration becomes unsuccessful. 

Considering the above points, we study misbehavior detection using the local voting game in the presence of uncertainty. Our main contributions in this paper can be summarized as follows:


\begin{itemize}[leftmargin=*]

\item We develop one stage of a local voting game using a plain Bayesian game. We capture the uncertainties of a node w.r.t. its detection system, the type of the target node, and strategies of other players in the game. In addition, we consider incentives in expected utilities (payoffs) of players to encourage nodes to cooperate.



\item We analyze the proposed model using a mixed-strategy BNE to obtain the equilibrium points of the game. Our findings reveal the best strategies that can be adopted by attackers and benign players w.r.t. the game parameters. Specifically, we ensure that no node can improve its utility by changing its strategy.

\item We provide extensive numerical results to verify the analysis and investigate the impact of cooperation parameters on the identification of malicious nodes. Our results confirm the influence of the designed incentives, hence participation rate, on the strategies of malicious and benign nodes. We observe, in particular, that if the participation incentives go beyond a certain limit, then \emph{correct} target-node identification will be decreased, in spite of the growing participation rate.

\end{itemize}

The remainder of this paper is organized as follows. Section \ref{sec:assum} describes assumptions, the local voting game, and the objectives of this paper. Section \ref{sec:prob_formu} formulates the game that includes defining parameters, payoff design, and a variable benefit scheme in the game. Section \ref{sec:BNE} applies Bayesian game analysis to derive equilibrium points in the proposed model. Section \ref{sec:results} is devoted to the numerical results. Section \ref{sec:conc} concludes the paper.


\vspace{-0.15 in}


\section{ Assumptions and Problem Description}\label{sec:assum}

\subsection{Network Model}
We study misbehavior detection in a VANET where nodes have short-lived connections, and a centrally managed station is absent. 
We assume that nodes (i.e., vehicles) are powerful enough to have wireless communication among themselves. We also assume that nodes have the same range of communications.
We consider a contention-based medium, e.g., IEEE $802.11$p in a VANET, that can represent the sequential nature of wireless channel access \cite{Raya08revoc}. We further assume that a base station or a certificate authority has already established the credential of nodes, hence each node has a unique ID. 

We presume that there are two types of nodes in the network: malicious and benign. Malicious nodes may attack benign nodes by disseminating false information. For example, a malicious car might inject faulty data to the sensors of the car that follows it, in order to manipulate an optimal space between them \cite{fer18}. On the other hand, a benign node is equipped with a monitoring system to detect abnormal or counterfeit signals. 
For example, an autonomous vehicle can use a set of anti-spoofing techniques to detect fake GPS signals \cite{VTC18}.
 However, benign nodes do not necessarily need to monitor all of their neighbors due to the cost of monitoring over all short-lived connections.

\vspace{-0.15in}
\subsection{Local Voting Game}
We assume that nodes can participate in a local voting game in order to determine the identity of a node in the network. The voting game starts when an initiator broadcasts the ID of a target node. Then, neighboring nodes choose either to vote or not to vote (abstain) on the type of the target node. 
Each node calculates its costs and benefits to choose a strategy.
The nodes broadcast their decisions sequentially, and each node's decision is made in one stage of the game.
We assume that the belief of a node w.r.t. the target node is independently inferred and does not change (e.g., by other votes) during the game. We presume that the target node is identified when the number of votes in one type (either malicious or benign) reaches a pre-defined number. This number is denoted by $n_{th}$. If correct (wrong) votes reach $n_{th}$, then we will have correct (wrong) target node identification.
If $n_{th}$ is not reached during the game, then we will have \emph{undecided} target node identification.

Malicious nodes and benign nodes can choose some strategies in the game.
A malicious node could select to attack or not to attack a benign node. 
On the other hand, a benign node might or might not use its detection system to monitor its neighbors. After a target node is determined, a benign node checks whether it has already monitored the target node. If it has not monitored the target node, then it will abstain from voting, simply because it does not have any information about the node. But, if the benign node has monitored the target node, then it calculates its payoffs. If its voting payoff outweighs its abstaining payoff, then the benign node will vote; otherwise, it will abstain. On the other hand, malicious nodes always vote against a benign target node and for a malicious target node. We do not consider strategic malicious nodes that can optimize their types of votes to collect some credits, or send multiple wrong votes (Sybil attack \cite{yu13det}).

\vspace{-0.15 in}
\subsection{Problem Definition}

We assume that malicious nodes are aware of an existing voting game in the network. The objective of a malicious node is to maximize the level of its aggressiveness in the network without being identified. However, it is uncertain about the probability of being monitored by a benign node, the accuracy of a monitoring system, and the strategy of a benign node in the game (i.e., voting or abstaining). In contrast, a benign node knows that some of its neighbors may be malicious.
The objective of a benign node is to choose a strategy with the aim of target node identification. However, a benign node has some limitations in its monitoring system. Also, it is uncertain about the strategies of malicious nodes. Therefore, it is uncertain about the type of the target node.
Taking these points into consideration, our goal is the following:
\begin{itemize}[leftmargin=*]
\item To design payoffs for a benign node w.r.t. the explained uncertainties and the value of its contribution in the game,
\item To determine the best strategies for malicious nodes and benign nodes.
\end{itemize}

We address the first problem in section IV by considering the following: (i) the vote of a benign node that could be either correct or incorrect; (ii) the probability of correct target node identification in each stage, which is mainly based on the votes that have already been cast;  and (iii) the impact of a benign node's strategy on correct, wrong, and undecided target node identification. We address the second problem in section V. In particular, we develop one stage of the voting game using a Bayesian game to study the reactions of a benign node w.r.t. a benign or malicious target node. This helps us understand the best strategies of both types of nodes in the network.

\vspace{-0.10 in}

\section{Problem Formulation} \label{sec:prob_formu}
In this section, we first define parameters of the game. Then, we focus on designing payoffs based on individual and group beliefs of players. 

\vspace{-0.15in}
\subsection{Parameters}\label{Parameters}

For our analysis, we need to define some parameters, as listed in Table \ref{symbol}. To begin, we assume that a benign node holds an asset with a security value of $w$, where $w>0$. A malicious node could compromise the asset by paying the cost of an attack, denoted by $c_a$. In contrast, a benign node protects its asset by monitoring for attacks, with probability $P_m$. This monitoring costs $c_m$ for the node, and all costs are positive. It is sensible to assume that $w>c_a$ and $w>c_m$. Otherwise, the attacker and the benign node lose their motivations to attack and protect the asset, respectively. A benign node assigns a prior probability of $\mu$ for its neighbors to be malicious.
The monitoring system of a benign node can detect an abnormality with probability $\alpha$ (i.e., true positive rate), while it suffers from a false alarm (i.e., false positive rate) with probability $\beta$. It is rational to expect that $\alpha>0.5>\beta$.

\begin{table}[t]
\centering
\captionsetup{justification=centering}
\caption{List of parameters in alphabetical order.}
{\includegraphics[width = 3.5 in , height=3.8 in]{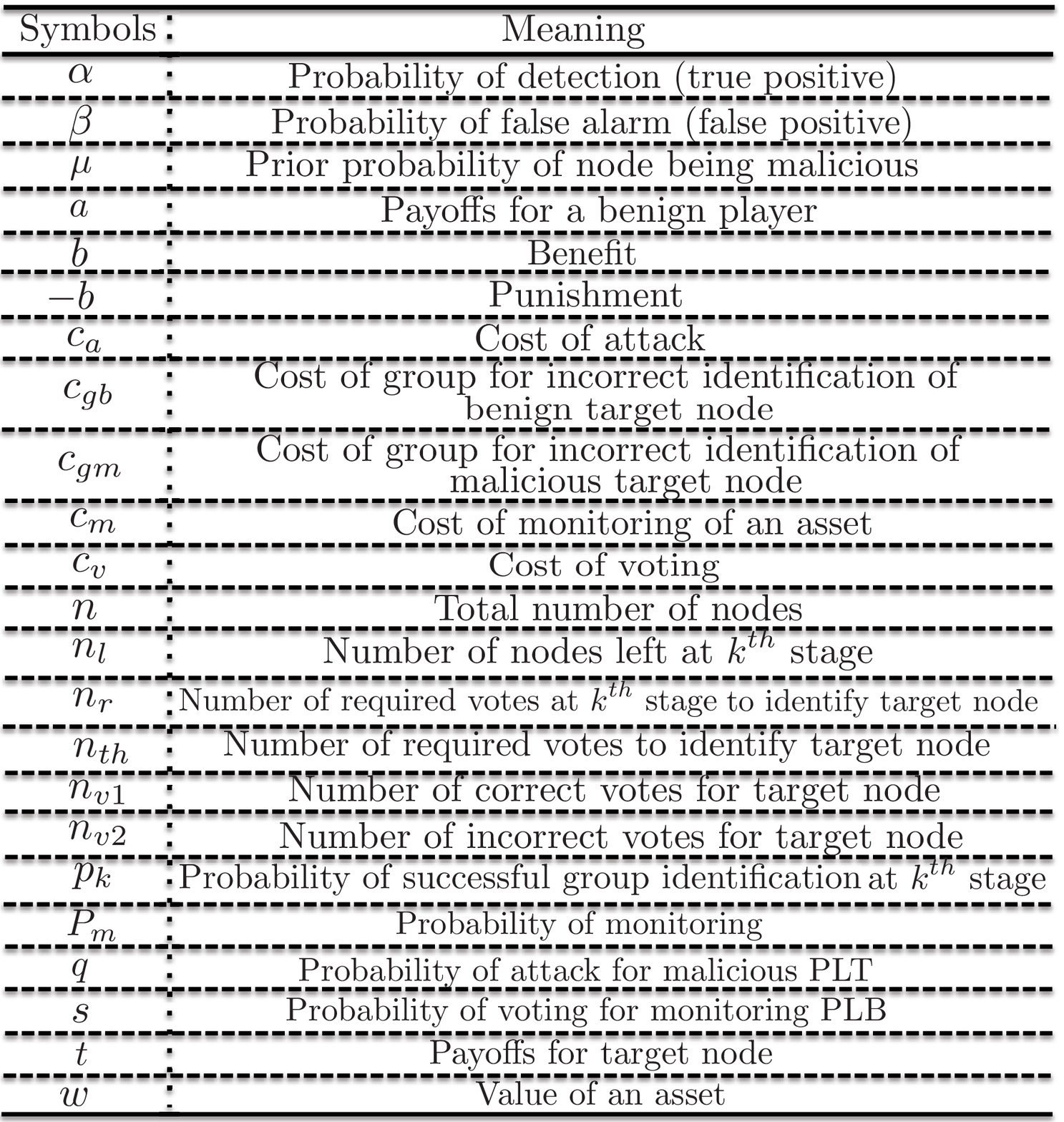}}\label{symbol}
\vspace{-0.35 in}
\end{table}

It is assumed that $n$ nodes are in a neighboring area.  Each benign node can vote by paying $c_v$ as the cost of voting. The benefit of a correct strategy and the punishment of an incorrect strategy for a benign node are denoted by $b$ and $-b$, respectively. It is assumed that $b>c_v>0$, which means that the benefit of a correct strategy (either voting or abstaining) is more than its cost.
To generalize the analysis, we design the game at the $k^{th}$ stage, in which the type of a target node has not yet been determined.
It is assumed that $n_{v1}$ correct votes and $n_{v2}$ wrong votes have already been cast before the $k^{th}$ stage of the game. In this stage, there are $n_l$ nodes left in the game. 
We let $n_r$ denote the number of remaining votes required to identify the target node.
 We use $p_k$ to denote probability of correct target node identification at the $k^{th}$ stage.
It is assumed that the cost of the group (neighboring nodes) for the incorrect identification of a malicious target node and a benign target node are $c_{gm}$ and $c_{gb}$, respectively. Equipped with these parameters, we design the expected payoffs for players in the game.




\vspace{-0.15 in}
\subsection{Payoff Design} \label{sec:payoffs}

\begin{figure}[t] 
\centering
\captionsetup{justification=centering}
{\includegraphics[width = 3.5in , height=1.4 in]{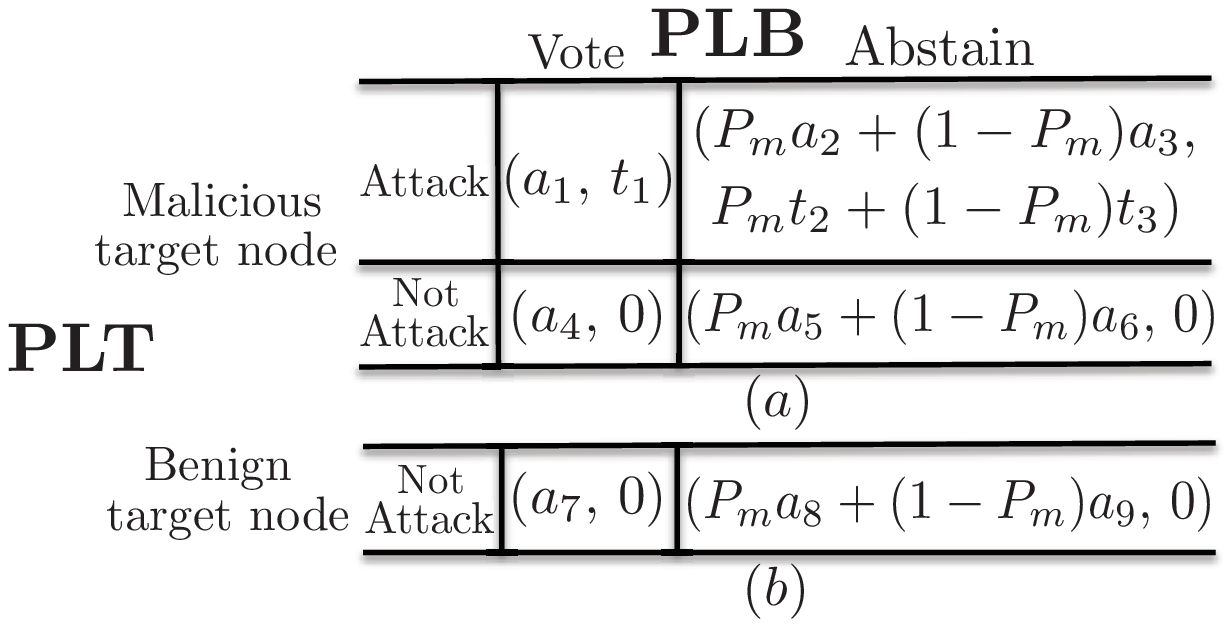}}
\caption{Players' Payoffs in the game relative to (a) malicious target node, and (b) benign target node.}
\vspace{-0.2 in}
\label{fig:Per_n}
\end{figure}

In this section, we study players' payoffs at the $k^{th}$ stage of the game, where the target node has not yet been identified. Fig. \ref{fig:Per_n} shows payoffs in the game, where rows and columns indicate the strategies of a target node and a benign player, respectively. Hereafter, the target node and benign player are denoted by PLT and PLB, respectively. The first element in each window refers to the PLB and the second element refers to the PLT. Here, $a_z$s refer to payoffs for PLB, and $t_z$s refer to payoffs for PLT, where $1 \le z \le 9$. It is assumed that $t_z = 0$ for $4 \le z \le 9$, because a non-attacking PLT does not gain or lose in the game. We define each player's payoff as the summation of an individual payoff and a group payoff. That is, $a_z = a_{z,i} + a_{z,g}$ and $t_z = t_{z,i} + t_{z,g}$, where $a_{z,i}$ and $t_{z,i}$ denote individual payoffs, and $a_{z,g}$ and $t_{z,g}$ denote group payoffs. The individual payoff only considers interactions between two players. In contrast, the group payoff accounts for the impact of a player's strategy on all members in the neighborhood.

The PLT could have either attacked or not attacked a PLB, depending on its type and strategy. Three scenarios could have happened between PLT and PLB: (i) malicious PLT attacked PLB, (ii) malicious PLT did not attack PLB, and (iii) PLT is benign. The PLB at the $k^{th}$ stage of the game, however, chooses its voting or abstaining strategy based on what it has observed before $k^{th}$ stage and what it might achieve in the game. In what follows, we comprehensively study how to obtain payoffs in scenario I. Payoffs for scenarios II and III can be derived in the same fashion.


The first row in Fig.\ \ref{fig:Per_n}(a) pertains to scenario I (i.e., malicious PLT attacked PLB). Here, we are interested in obtaining $a_z$s and $t_z$s, where $1\leq z \leq3$. The subscripts $z=1$ and $z=2$ refer to the payoffs of a monitoring PLB, and $z=3$ refers to the payoffs of a non-monitoring PLB.
To obtain $a_{1,i}$ and $a_{2,i}$, note that a monitoring PLB pays $-c_m$ as the cost of monitoring. 
Also, a monitoring PLB gains $(2\alpha-1)w$ (i.e., $\alpha w-(1-\alpha)w$)) from its detection system, which includes the impact of detection rate ($\alpha$) and false negative rate ($1 - \alpha$). Thus, we have $a_{1,i} = a_{2,i} = -c_m + (2\alpha-1)w$. In addition, PLT's attack compromises a non-monitoring PLB's asset, i.e. $a_{3,i} = - w$. On the other hand, PLT pays $-c_a$ as the cost of the attack. If PLB is in a monitoring state, then the loss of PLT can be assumed as the negative gain of PLB's individual payoff \cite{Liu2006}, i.e. $-(2\alpha -1) w$. Hence, we have $t_{1,i} = t_{2,i} = -c_a - (2\alpha -1) w$. However, if PLB is in a non-monitoring state, then PLT gains $w$ from its attack, i.e., $t_{3,i}= -c_a + w$.

\begin{figure}[t]
\centering
\captionsetup{justification=centering}
{\includegraphics[width = 3.3 in , height=2 in]{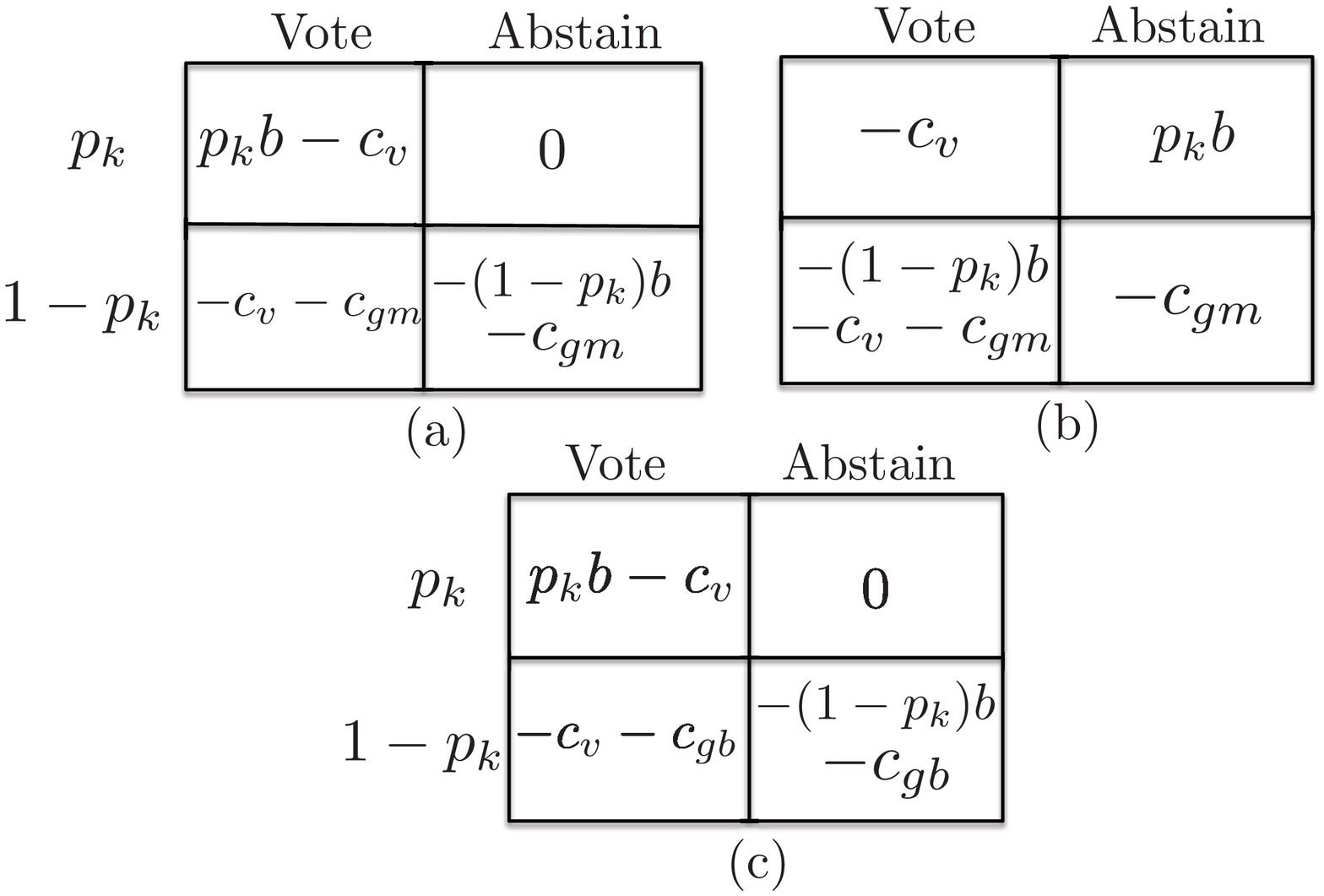}}
\caption{Group Payoffs for three scenarios: (a) malicious target node has attacked a monitoring benign node, (b) malicious target node has not attacked a monitoring benign node, (c) benign target node versus a monitoring benign node.}
\vspace{-.2 in}
\label{fig:benefits_n}
\end{figure}

To design $a_{2,g}$ and $a_{3,g}$, the voting payoff and the abstaining payoff of a monitoring PLB are studied w.r.t. the probability of correct target node identification ($p_k$). This is because the target node is not yet identified at the $k^{th}$ stage of the game, and hence correct, wrong, or undecided target node identification may happen. Fig. \ref{fig:benefits_n}(a) shows the strategies of a monitoring PLB at the $k^{th}$ stage relative to $p_k$. The left column corresponds to the player's voting payoff (i.e., $a_{1,g}$), and the right column refers to its abstaining payoff ( i.e., $a_{2,g}$). Here, $-c_v$ and $0$ represent the cost of voting and abstaining, respectively. Also, $-c_{gm}$ in the lower row denotes the cost of incorrect identification of a malicious target node. In addition, the reward of voting in correct target identification (top left window) and the punishment of abstaining in incorrect target identification (bottom right window) are represented by $bp_k$ and $-b(1-p_k)$, respectively. These are proportional to $p_k$ because the player's expected outcome is entangled with the probability of correct target node identification in the middle of the game. The reward and the punishment are considered (as incentives) to encourage nodes in cooperation. 

\begin{figure}[t] 
\centering
\captionsetup{justification=centering}
{\includegraphics[width = 3.2 in , height=1.4 in]{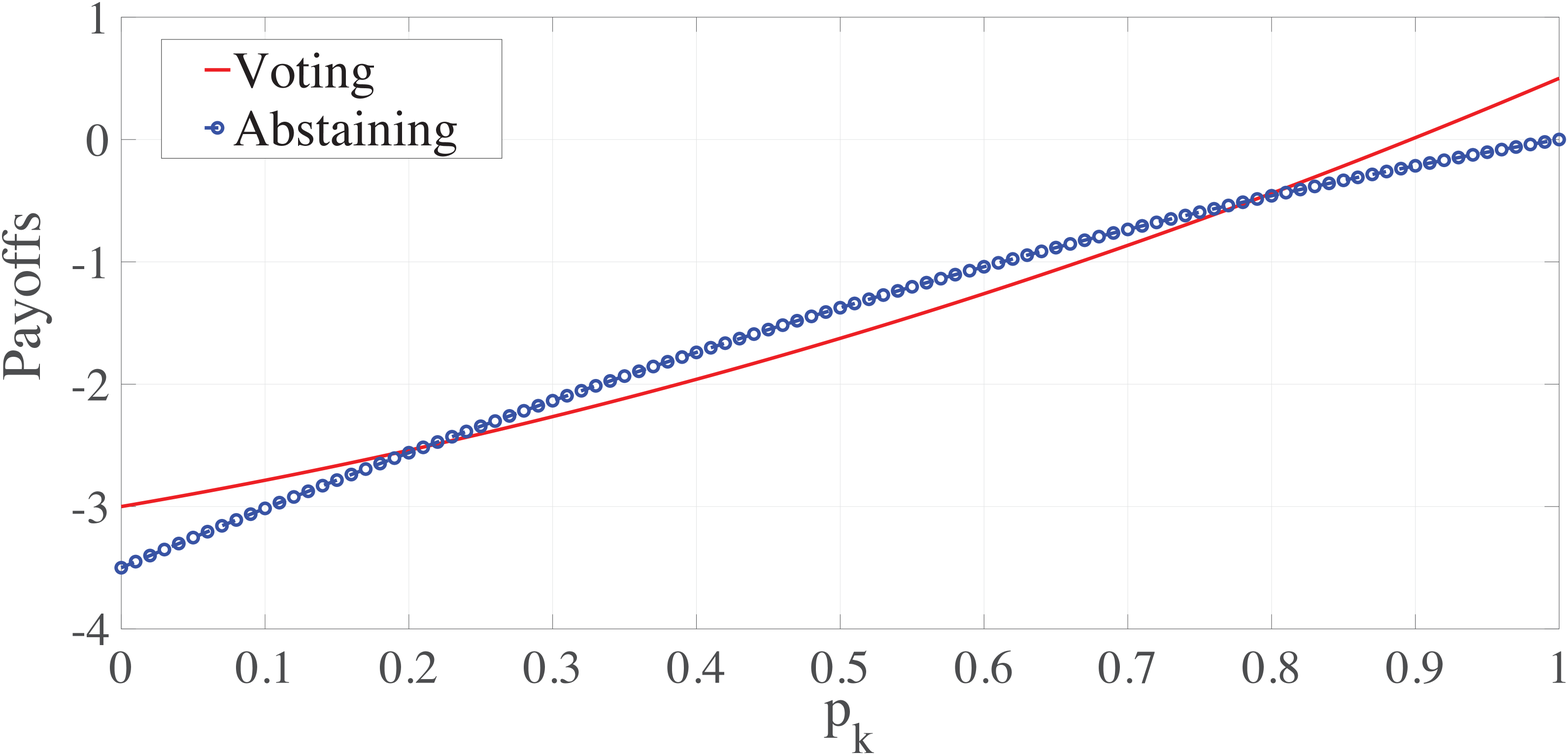}}
\caption{Payoffs for monitoring benign node relative to $p_k$.
}
\vspace{-0.25 in}
\label{fig:Pay_1}
\end{figure}

Fig. \ref{fig:Pay_1} shows payoffs for PLB for different $p_k$s, in which $c_v=1$, $b=1.5$, and $c_{gm}=2$. As can be seen, voting payoffs and abstaining payoffs outweigh each other, depending on the value of $p_k$. For instance, voting payoffs are dominant for $p_k<0.2$, and thereby the player votes in this interval.
In this case, voting can be interpreted as an attempt from the player to increase $p_k$ and avoid an incorrect outcome of the game. The main motivation of the player, however, comes from the punishment of the game. In other words,
the cost of voting is lower than the punishment of the game when the malicious target node is not correctly identified (lower row of Fig. \ref{fig:benefits_n}(a)). That is, if $p_k\rightarrow 0$, then $-c_v > -(1-p_k)b$.  Thus, the player votes not only to increase $p_k$ but also to avoid punishment.

Using above individual payoffs and group payoffs in Fig. \ref{fig:benefits_n}(a), we can compute payoffs in this scenario.

\begin{lem} $a_z$ and $t_z$, where $1\leq z \leq3$, are as follows:
\begin{align}
& a_1 = \, p_k^2b-c_v-(1-p_k)c_{gm} - c_m + (2\alpha-1)w, \label{eq:b1}  \\ 
& a_2 = -(1-p_k)^2b-(1-p_k)c_{gm} - c_m + (2\alpha-1)w, \label{eq:b2} \\
& a_3 = - w  - (1 - p_k)c_{gm}, \label{eq:b3} \\
& t_{1}  = t_{2} = - c_a - (2\alpha -1) w + (1-p_k)c_{gm}, \label{eq:t1} \\
& t_{3} = - c_a + w +  (1-p_k)c_{gm}. \label{eq:t3}
\end{align}
\label{lem:a1a2}
\vspace{-0.3 in}
\end{lem}
\noindent All lemmas and theorems are proved in the appendix.

For the second and third scenarios, we use Fig. \ref{fig:benefits_n}(b) and Fig. \ref{fig:benefits_n}(c) respectively to obtain group payoffs. Applying a similar reasoning of scenario I to scenarios II and III yields the rest of the payoffs as follows:
\begin{align}
& a_4 = -(1-p_k)^2b - c_v - (1-p_k)c_{gm} -c_m-\beta w, \label{eq:b4} \\
& a_5 = \, p_k^2b-(1-p_k)c_{gm} -c_m-\beta w, \label{eq:b5} \\
& a_6 = - (1 - p_k)c_{gm}, \label{eq:b6} \\
& a_7 = \, p_k^2b-c_v-(1-p_k)c_{gb}  -c_m-\beta w , \label{eq:b7} \\
& a_8 = -(1-p_k)^2b-(1-p_k)c_{gb} -c_m-\beta w, \label{eq:b8} \\
& a_9 = - (1 - p_k)c_{gb}. \label{eq:b9}
\end{align}

As seen in equations (\ref{eq:b1})-(\ref{eq:b9}), $p_k$ plays an important role in the payoffs. Thus, we obtain $p_k$ to evaluate the strategies of players. It is noteworthy that the value of $p_k$ increases when PLB votes correctly. This improvement in $p_k$ is denoted by $\delta$.

\begin{lem} $p_k$ and $\delta$ can be obtained as follows:
\begin{align}
p_k & = \sum_{i = n_r}^{n_l} \dbinom{n_l}{i} \ \big(p_s\big)^{i} \ \big( 1-p_s \big)^{n_l - i}, \label{eq:pk} \\
\delta & = \ \dbinom{n_l}{n_{r}-1} \ \big(p_s\big)^{n_{r}-1} \ \big( 1-p_s \big)^{n_l - (n_{r}-1)}, \label{eq:delta}
\end{align}
where $p_s$ represents the probability of correct target identification by remaining nodes in the game.
\label{lem:pk}
\end{lem}

\vspace{-0.20 in}
 
\section{Equilibrium Analysis} \label{sec:BNE}
\color{black}
The objective of the players is to maximize their payoffs in the game. In this regard, we obtain possible equilibrium points using a Bayesian game to better understand the behavior of the players. 
In particular, we obtain the best strategies of benign players to identify a malicious node, while we find the maximum level of aggressiveness for malicious nodes without being identified. 
 In this respect, we use the interactions between a PLB and a PLT, as illustrated in Fig. \ref{fig:Per_n}. \color{black}
To obtain equilibrium points, we use a mixed-strategy BNE because the game is a finite strategic-form game.
To determine each player's indifference strategy, we define $q$ as the probability of attack for a malicious PLT, and $s$ as the probability of voting for a monitoring PLB.

\begin{theo} 
Given $\mu$ and $P_m$, the game defined in section III has a mixed-strategy BNE, which is as follows:
 \begin{itemize}
\item Malicious node attacks with a probability of $q^{*}$, which is
\begin{small}
\begin{align}
q^{*} = \frac{ q_1 + q_2 + ... + q_n}{n}, \label{eq:q_star} 
\end{align}
\end{small}
where $q_k$, is the probability of attack for the $k^{th}$ node
\begin{small}
\begin{align}
q_k & = \frac{ A_k}{B_k}, \label{eq:q_k}
\end{align}
\begin{align}
\notag A_k &= \mu \big(1+P_m\big) \big( 2 p_k^{2} - 2p_k + 1\big)b +  \big(1 - P_m\big)  \\
\notag &\quad \times \big(c_m + \beta w\big) + c_v -p_k^2 b - P_m \big( 1- p_k\big)^2 b, \\
\notag B_k &=  \mu \big(1+P_m\big) \big( 2 p_k^{2} - 2p_k + 1\big)b + \mu \big(1-P_m\big)(2\alpha + \beta)w.
\end{align}
\end{small}
\item Monitoring benign node votes with probability of $s^{*}$, which is equal to
\begin{small}
\begin{flalign}
s^{*} =  \frac{c_a+(2\alpha P_m - 1)w -c_{gm}}{(1-P_m)[-c_a+ (1-2\alpha)w +c_{gm}] - \delta c_{gm}}. \label{eq:s_star}
\end{flalign}
\end{small}
\end{itemize}\label{theo:mixed_BNE}
\end{theo}
Note that the mixed-strategy provides general equilibrium points w.r.t. different parameters. 
In a special case, if all nodes monitor their neighbors, i.e. $P_m=1$, then one can derive an upper bound for the benefit and a lower bound for the detection rate using eqs. (\ref{eq:q_k}) and (\ref{eq:s_star}), respectively.

\begin{cor} In Theorem $2$, if $P_m=1$, then we have
\begin{small}
\begin{align}
b < \frac{c_v}{(1-2\mu)(2p_k^{2}-2p_k+1)}, \label{eq:b_bound}
\end{align}
\begin{align}
 \alpha > \frac{w-c_a+c_{gm}(1-\delta)}{2w}. \label{eq:alpha_bound}
\end{align}
\end{small}
\end{cor}
\noindent From eq. (\ref{eq:b_bound}), we observe that as $\mu \rightarrow \frac{1}{2}$, the upper bound increases. This allows network designers to select higher values of benefit in an environments where the probability of malicious PLT is higher. On the other hand, eq. (\ref{eq:alpha_bound}) implies that a monitoring system must have a minimum true positive rate in order to make a malicious node indifferent in the game.


\vspace{-0.1 in}

\section{Numerical Results} \label{sec:results}

To evaluate our analysis, we assume that 40 nodes run the game in an area of $625$ m $\times$ $625$ m (normal density $\approx 100 \frac{\text{nodes}}{km^{2}}$ in \cite{density16}). Since the analysis is probabilistic, we run 100 iterations for each simulation. Then, we take an average of the results with $95\%$ confidence interval. The default game parameters are as follows:
\begin{itemize}
\item Monitoring system parameters: $\alpha = 0.95$, $\beta = 0.05$,
\item Probabilities: $P_m=0.75$, $\mu =0.2$, and $q =0.4$,
\item Costs and benefits: $c_{gb} = c_{gm} = 4 $, $w = 4 > b=3 > c_m = c_a = c_v = 1$.
\end{itemize}
If we change these parameters to better explain a scenario, then we will explicitly mention it. We describe the results in three subsections. Initially, we study the impact of incentives (in particular, $b$) on correct, wrong, and undecided target node identification. Then, we focus on the behavior of malicious nodes w.r.t. their portion and aggressiveness in the network. Finally, we compare our work with scenarios where the uncertainties discussed in this paper have not been considered, e.g., \cite{masdari17, bil10opt, Abass17evol}.

\begin{figure}[t] 
\centering
\captionsetup{justification=centering}
{\includegraphics[width = 3.45 in , height = 1.5 in]{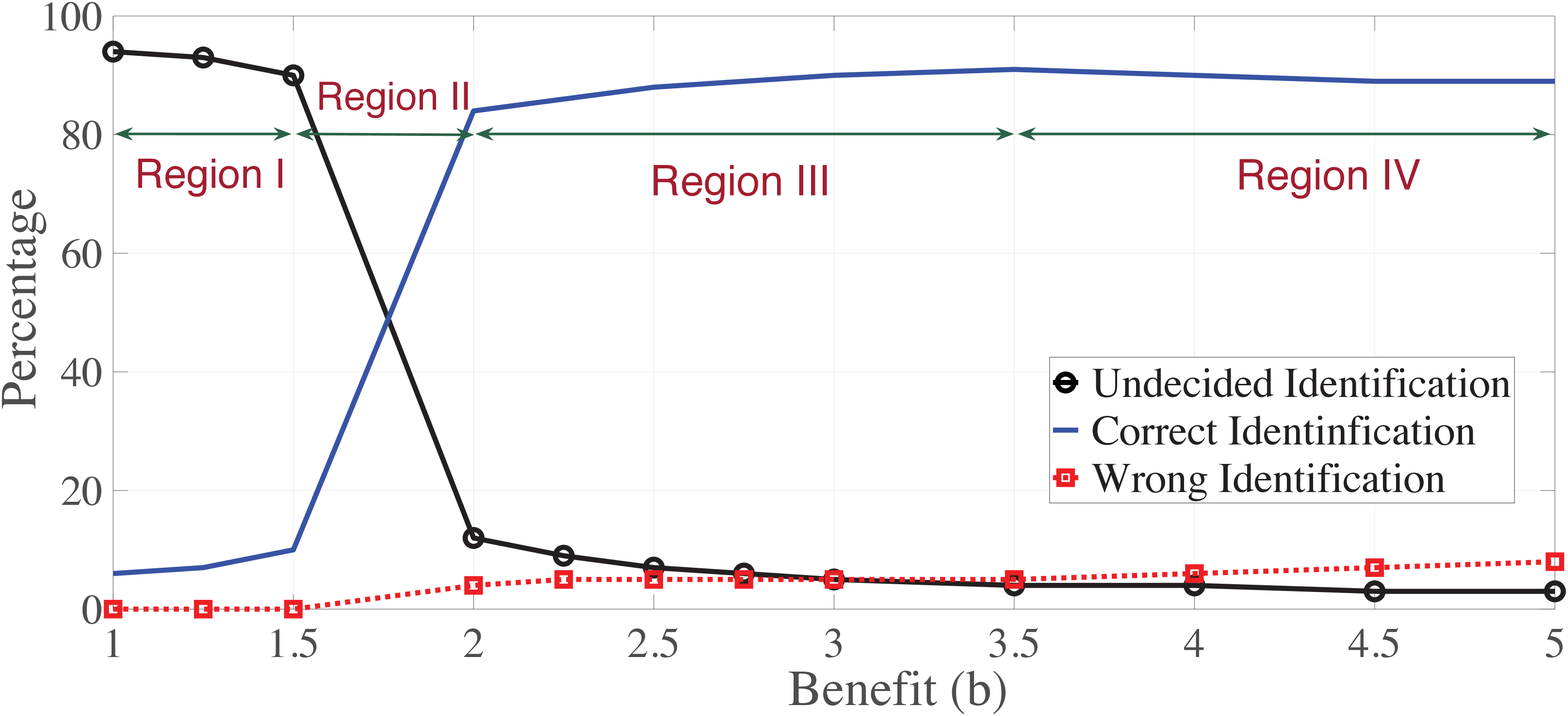}}
\caption{ Game outcomes versus variation of benefits.}
\vspace{-0.15 in}
\label{fig:Ben}
\end{figure}

\vspace{-0.15 in}
\subsection{Impact of incentives}

Fig. \ref{fig:Ben} illustrates the percentage of target node identification versus $b$. Here, it is assumed that $q=0.7$.
As shown, this figure can be categorized into four different regions. In region I, the percentage of undecided target identification outweighs correct and wrong identifications for a simple reason: the benefit is not large enough to persuade nodes to participate in the game.
Region II, however, illustrates a drastic reduction of undecided identification. This indicates that voting payoffs become larger in comparison to abstaining payoffs.
In addition, correct identification dominates over wrong identification, which is the result of the following: (i) benign nodes with high monitoring and detection rates (i.e., $P_m=0.75$ and $\alpha = 0.95$), and (ii) malicious nodes with a high level of aggressiveness (i.e.,  $q=0.7$).
Region III shows a slight increase in correct identification and a decrease in undecided identification because of lower payoffs for abstaining from the game. The increase of wrong identification over undecided identification is remarkable in region IV. Wrong votes in this region mainly come from highly encouraged benign nodes that have not been attacked by a malicious target node. In other words, since voting payoffs are significantly larger than abstaining payoffs (i.e., $a_4 > a_5$ and $a_7 > a_8$), a benign node votes in favor of a non-attacking target node.
This observation reveals that persuading every node to vote by applying the leverage of benefit does not necessarily lead to a better outcome. Taking all regions into consideration, region III indicates the best option for the benefit design.

\begin{figure}[t] 
\centering
\captionsetup{justification=centering}
{\includegraphics[width = 3.5 in , height = 1.5 in]{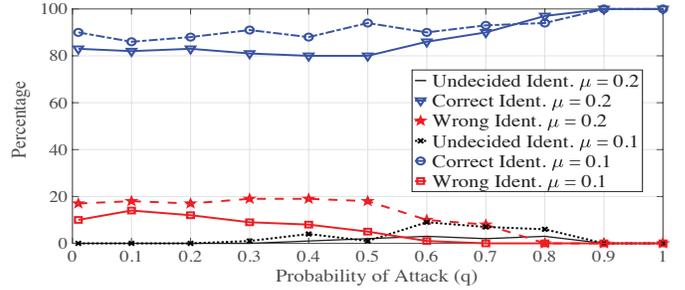}}
\caption{ Impact of portion of malicious nodes ($\mu$) and probability of attack ($q$) on identification results.}
\vspace{-0.2 in}
\label{fig:Qs}
\end{figure} 

\vspace{-0.18 in}
\subsection{Impact of malicious nodes}
Fig. \ref{fig:Qs} shows the percentage of target identification w.r.t. the portion of malicious nodes and their probability of attack ($q$) in the network. As shown, when $q$ increases, correct identification generally increases, which confirms that aggressive attackers can be identified easier. However, wrong identification is reduced after a certain value of $q$; for example, $q = 0.1$ for $\mu = 0.1$. When the number of malicious nodes increases in the network, this decreasing trend starts at higher values of $q$; for instance, $q = 0.4$ for $\mu = 0.2$. This reveals that malicious nodes become more aggressive when their number increases.

\vspace{-0.15 in}
\subsection{Comparison}
In this section, we compare our work with the scenarios where some of the explained uncertainties have not been considered (e.g.\cite{masdari17, bil10opt, Abass17evol}). It is worth mentioning that the comparison is limited to highlighting uncertainties in those scenarios. This is because the nature of their games and objectives are slightly different. However, this comparison provides us with insights into the impact of imperfect information at nodes on the outcome of a local voting game.

\begin{figure}[t] 
\centering
\captionsetup{justification=centering}
{\includegraphics[width = 3.5 in , height = 1.45 in]{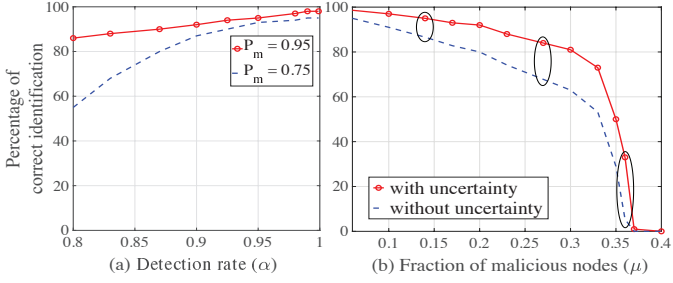}}
\caption{\small Impact of uncertainties in game in relation to: (a) detection rate, and (b) correct identification rate.}
\vspace{-0.22 in}
\label{fig:comp}
\end{figure}

Fig. \ref{fig:comp}(a) shows the impact of the true positive detection rate ($\alpha$) on the correct target identification. As can be seen, it is essential for nodes to have high values of $\alpha$ to gain high correct target identification. The value of $\alpha$ becomes more important when fewer benign nodes monitor their neighbors (i.e., smaller $P_m$).
Fig. \ref{fig:comp}(b) indicates a comparison between a design with and without uncertainties in the local voting game. In particular, we assume that a design without uncertainty has the following parameters: $\alpha = 1, \beta = 0$, and $q = 1$.
As shown, the difference between graphs is growing with $\mu$. This is because a player without uncertainty considers a non-attacking malicious node as a benign node and votes for it.
Our proposed design, on the other hand, prevents benign nodes from voting when they are unsure about the strategy of malicious nodes. 
Moreover, in both cases, when $\mu$ goes beyond a threshold, here $0.3$, correct  identification is significantly reduced. This comes from higher payoffs for abstaining in comparison to voting. Interpreted differently, benign nodes are unwilling to cooperate in a game that a high portion of participants are malicious.

\vspace{-0.27 in}
\section{Conclusion}\label{sec:conc}
In this paper, we have provided a game-theoretic approach to identify malicious nodes in VANETs, where central stations are not available. In particular, we have studied the strategies of nodes in a local voting-based game using a Bayesian game, in which nodes have incomplete information about the accuracy of their monitoring systems, the type of neighbors (benign or malicious), and the outcome of the game. By offering incentives in expected utilities, we have provided encouragements for game participation with the aim of improving correct node identification. We have derived a mixed-strategy BNE points to study the best strategies of players in the game. Simulation results showed the impact of different parameters such as participation benefits and detection rate on identification of malicious nodes. 

\vspace{-0.20 in}
\section{ Appendix}
%
%

\vspace{-0.05 in}
\subsection{Proof of Lemma \ref{lem:a1a2}}
\begin{proof}
In Fig. \ref{fig:benefits_n}(a), the left column (vote) refers to $a_{1,g}$, and the right column (abstain) refers to $a_{2,g}$. Since each strategy (voting or abstaining) has two identification possibilities, denoted by $p_k$ and $1-p_k$, the payoff of each strategy should be weighted by corresponding probabilities. In other words,
\begin{align}
\notag a_{1,g} & =  p_k \times \big( p_kb - c_v\big) \, + \big(1-p_k\big) \times \big(- c_v  - c_{gm}\big), \\
\Rightarrow a_{1,g} & =  p_k^2b-c_v-(1-p_k)c_{gm}, \label{eq:b1g} \\
\notag a_{2,g} & =  p_k \times \big( 0 \big) \, + \big(1-p_k\big) \times \big(- (1-p_k)b - c_{gm}\big), \\
\Rightarrow a_{2,g} & = -(1-p_k)^2b-(1-p_k)c_{gm}. \label{eq:b2g}
\end{align}

If we add individual payoffs $a_{1,i} = a_{2,i} = - c_m + (2\alpha-1)w$ to eqs. (\ref{eq:b1g}) and (\ref{eq:b2g}), we obtain eqs. (\ref{eq:b1}) and (\ref{eq:b2}), respectively. Eq. (\ref{eq:b3}) can be obtained by adding $a_{3,i}$ to $a_{3,g}$. To have $a_{3,g}$, we know that non-monitoring PLB abstains from the game, so it completely relies on other nodes for the group payoff. Thus, we define $a_{3,g} = - (1 - p_k)c_{gm}$, where the node does not impact the group decision. If $p_k=1$, then the node is not harmed, but if $p_k=0$, then it gets $-c_{gm}$. The summation of $a_{3,i}$ and $a_{3,g}$ yields eq. (\ref{eq:b3}).

On the other hand, to obtain PLT's payoffs, we need to have PLT's group payoff. $t_{k,g}$s for $1 \le k \le 3$ can be expressed as $(1 - p_k) c_{gm}$, which reflects the inverse proportional relationship between $p_k$ and the gain of malicious PLT. The summation of individual payoffs and $(1 - p_k) c_{gm}$ ( as group payoff) yields eqs. (\ref{eq:t1}) and(\ref{eq:t3})
\end{proof}

\vspace{-0.25 in}
\subsection{Proof of Lemma \ref{lem:pk}}
\begin{proof}
To obtain $p_k$, note that $n_{v1}$ and $n_{v2}$ votes have already been cast until the $k^{th}$ stage, while there are $n_l$ nodes left in the game. To derive a closed form for $p_k$, note the following: (i) If $n_r$ > $n_l$, then $p_k=0$, which means that the number of left nodes is less than the number of required votes to identify PLT; (ii) if $n_r = 0$, then $p_k=1$, which implies that PLT has been already identified; (iii) $p_k$ directly depends on $n_l$ and their type; and (iv) if $n_{r}$ is reduced, then $p_k$ will be increased. Taking these points into account, $p_k$ can be written in the form of eq. (\ref{eq:pk}), where $p_s$ represents the probability of correct target node identification. For instance, assume $n =10$, $k =7$ (i.e., $n_l=3$), $p_s=1/3$, and $n_{th}=4$. Under such assumptions, if $n_{v1}=0$ (i.e., $n_r =4$), then equation (\ref{eq:pk}) yields $p_k=0$ because of the first condition. If $n_{v1}=4$ (i.e., $n_r=0$), then eq. (\ref{eq:pk}) yields $p_k =1$ because of the second condition. Also, substituting  $n_r=1$ and $n_r =3$, respectively, yields $p_k = 0.7$ and $p_k \approx 0.04$, which confirm the last condition.  It is noteworthy that $p_s \propto \lambda(1-\mu)\alpha P_m$, where $\lambda$ represents the probability of remaining nodes to be in the network. 

Since $\delta$ is defined as the difference that a correct vote can make in $p_k$, we have \small{$\delta = p_k(voting)-p_k(abstaining)$}. That is
\begin{small}
\begin{align}
\notag  \Rightarrow \ \ & \delta =  \sum_{i = n_r-1}^{n_l} \dbinom{n_l}{i} \ \big(p_s\big)^{i} \ \big( 1-p_s \big)^{n_l - i} \ -  \\
 & \qquad \ \  \sum_{i = n_r}^{n_l} \dbinom{n_l}{i} \ \big(p_s\big)^{i} \ \big( 1-p_s \big)^{n_l - i},
\end{align}
\end{small}
which yields eq. (\ref{eq:delta}).
\end{proof}

\vspace{-0.25 in}

\subsection{Proof of Theorem \ref{theo:mixed_BNE}}
\begin{proof}
To obtain $q^{*}$, we first equalize the expected utilities for voting and abstaining to obtain $q_k$. Then, we take an average over all possible values of the $p_k$s to get eq. (\ref{eq:q_star}). In this way, we have the followings:
\begin{small}
\begin{align}
Eu \, \big[voting\big] & = Eu \, \big[abstaining\big]  \label{eq:vot_abs}
\end{align}
\end{small}
where,
\begin{align}
Eu \, \big[voting\big] & = \mu q a_1 + \mu \big(1- q \big) a_4 +  \big(1- \mu \big) a_7, \label{eq:Eu_vot} \\
\notag Eu \big[abstaining\big] & = \mu \, q \,P_m a_2 +  \mu \, q \big(1- P_m \big)a_3 + \mu \, \big( \, 1 -\\
\notag & \ \,  q \big) P_m a_5 +  \mu \big(1- q \big) \big(1- P_m \big) a_6 + \big(1 - \\
& \   \mu \big) P_m a_8 + \big(1- \mu \big) \big(1- P_m \big) a_9. \label{eq:Eu_abs}
\end{align}
Substituting eqs. (\ref{eq:b1}), (\ref{eq:b4}), and (\ref{eq:b7}) into eq. (\ref{eq:Eu_vot}), and eqs. (\ref{eq:b2}), (\ref{eq:b3}), (\ref{eq:b5}), (\ref{eq:b6}), (\ref{eq:b8}), and (\ref{eq:b9}) into eq. (\ref{eq:Eu_abs}), and then substituting eqs. (\ref{eq:Eu_vot}) and (\ref{eq:Eu_abs}) in eq. (\ref{eq:vot_abs}) yields eq. (\ref{eq:q_k}). Since the malicious PLT might attack the neighboring nodes regardless of their stage in the game, we take an average over all values of $q_k$s, which yields eq. (\ref{eq:q_star}).

To calculate $s^{*}$, we can equalize the expected utilities of attack and not attack from PLT, hence, obtaining
\begin{small}
\begin{align}
\mu \, s \, t_1 + P_m \, \mu \, (1-s) \, t_2 + (1- P_m) \, \mu \, t_3 = 0. \label{eq:s_prov}
\end{align}
\end{small}
Plugging eqs. (\ref{eq:t1}) and (\ref{eq:t3}) back into eq. (\ref{eq:s_prov}) yields eq. (\ref{eq:s_star}).
\end{proof}

\vspace{-0.35 in}

\linespread{0.965}
\bibliographystyle{ieeetran}
\bibliography{biblio1}
\IEEEpeerreviewmaketitle

\end{document}